\documentclass[ALICE,manyauthors]{cernphprep}

\usepackage[comma,square,numbers,sort&compress]{natbib}
\usepackage{hyperref}
\usepackage{lineno}
\usepackage{bm}

\begin{document}%

\begin{titlepage}
\PHyear{2018}
\PHnumber{107}      
\PHdate{11 May}  
%

\title{Inclusive J/$\psi$ production \\ in Xe--Xe collisions at \boldmath${\sqrt{{\textbf{\textit{s}}}_{\rm \bf{NN}}}}$ = 5.44~TeV}
\ShortTitle{Inclusive J/$\psi$ production in Xe--Xe collisions at ${\sqrt{s_{{\rm NN}}} = 5.44}$~TeV}   

\Collaboration{ALICE Collaboration\thanks{See Appendix~\ref{app:collab} for the list of collaboration members}}
\ShortAuthor{ALICE Collaboration} 

\begin{abstract}
Inclusive J/$\psi$ production is studied in \mbox{Xe--Xe} interactions at a centre-of-mass energy per nucleon pair of $\sqrt{s_{\rm NN}}= 5.44$ TeV, using the ALICE detector at the CERN LHC. 
The J/$\psi$ meson is reconstructed via its decay into a muon pair, in the centre-of-mass rapidity interval $2.5<y<4$ and down to zero transverse momentum. In this Letter, the nuclear modification factors $R_{\rm AA}$ for inclusive J/$\psi$, measured in the centrality range 0--90\% as well as in the  centrality intervals 0--20\% and 20--90\% are presented. The $R_{\rm AA}$ values are compared to previously published results for \mbox{Pb--Pb} collisions at 
$\sqrt{s_{\rm NN}}= 5.02$ TeV and to the calculation of a transport model. A good agreement is found between \mbox{Xe--Xe} and \mbox{Pb--Pb} results as well as between data and the model.

\end{abstract}
\end{titlepage}
\setcounter{page}{2}

%
%

The study of the production of quarkonium states plays an important role in the 
characterization of the properties of the Quark-Gluon Plasma (QGP)~\cite{Shuryak:1978ij}. This state of matter, where quarks and gluons are not confined into hadrons, can be produced in heavy-ion collisions at ultrarelativistic energies. 
Quarkonia are bound states of heavy quark-antiquark pairs (charmonia, c${\overline{\rm c}}$ and bottomonia, b${\overline{\rm b}}$) and their production rate is significantly affected by the QGP. In particular, the color force responsible for the binding of heavy quarks is expected to be screened in the QGP, leading to a suppression of quarkonium production which can be related to the initial temperature of the system~\cite{Matsui:1986dk,Digal:2001ue}. In addition, at very high energies, such as those available at the LHC, the abundant production of charm-anticharm pairs leads to a recombination process,  which may occur both in the QGP phase or when the system cools down and hadrons are formed out of the free quarks and gluons~\cite{BraunMunzinger:2000px,Thews:2000rj}. The study of the interplay between suppression and recombination processes offers the possibility of a quantitative investigation of the existence of colorless bound states of heavy quarks in the QGP.

An extended set of results was obtained for the J/$\psi$, a charmonium state with quantum numbers $J^{\rm {PC}}=1^{--}$, at LHC energies ($\sqrt{s_{\rm NN}}=2.76$ and 5.02 TeV) in \mbox{Pb--Pb} collisions~\cite{Abelev:2012rv,Chatrchyan:2012np,Abelev:2013ila,Adam:2015isa,Adam:2016rdg,Khachatryan:2016ypw,Sirunyan:2017isk}. Comparison of these results to theoretical  models~\cite{Zhao:2011cv,Du:2015wha,Andronic:2012dm,Ferreiro:2014bia,Zhou:2014kka} and to lower energy data~\cite{Adare:2011yf,Abelev:2009qaa} favors the picture described above. The study of the collision of nuclei lighter than Pb may give additional important information on the relative contribution of suppression and recombination mechanisms.

A step in this direction is performed in this Letter, where first results on J/$\psi$ production at LHC energies in \mbox{Xe--Xe}, a collision system ($A_{\rm Xe}=129$) lighter  than \mbox{Pb--Pb} ($A_{\rm Pb}=208$), are presented. Data were collected by the ALICE Collaboration at the centre-of-mass energy per nucleon pair $\sqrt{s_{\rm NN}}=5.44$ TeV, during a short run carried out at the end of 2017.
Due to the limited integrated luminosity, $L_{\rm int}\sim 0.34$~$\mu$b$^{-1}$, the statistical uncertainties are significantly larger than those of the  \mbox{Pb--Pb} results~\cite{Adam:2016rdg}, but nevertheless allow a meaningful comparison between the two systems, in terms of the nuclear modification factor $R_{\rm AA}$. This quantity is obtained as the ratio between the production yields in  nucleus--nucleus collisions and the corresponding proton--proton (pp) cross section, normalized to the nuclear thickness function $\langle T_{\rm AA}\rangle$~\cite{Miller:2007ri}. Values of $R_{\rm AA}$ smaller (larger) than unity indicate suppression (enhancement) effects for the particle under study. The results shown in this Letter correspond to the centre-of-mass rapidity range $2.5<y<4$, are integrated over transverse momentum ($p_{\rm T}$) and were obtained by studying the J/$\psi\rightarrow\mu^+\mu^-$ decay channel. The nuclear modification factor is studied as a function of the centrality of the collision~\cite{Abelev:2013qoq}, expressed as a percentage of the hadronic \mbox{Xe--Xe} cross section. 
The results correspond to inclusive J/$\psi$ production, which is the sum of a prompt component (directly produced J/$\psi$ and feed-down from other charmonium states) and a non-prompt component, due to the decay of particles containing a b quark. 

ALICE is the LHC experiment dedicated to the study of nuclear collisions, and is  described in detail in Refs.~\cite{Aamodt:2008zz,Abelev:2014ffa}. The main detector used in this analysis is a muon spectrometer~\cite{Aamodt:2011gj}, covering the pseudorapidity range $-4<\eta<-2.5$~\footnote{In the ALICE reference frame, the muon spectrometer covers a negative $\eta$ range and consequently a negative $y$ range. We have chosen to present our results with a positive $y$ notation.}. It includes tracking and trigger chambers, and reconstructs muons with $p_{\rm T}$ larger than a given threshold, which is set at the trigger level. In addition, the V0~\cite{Abbas:2013taa}, a set of scintillator detectors covering $2.8<\eta<5.1$ and $-3.7<\eta<-1.7$, is used to define the minimum bias (MB) interaction trigger via a coincidence of signals at positive and negative $\eta$ values. The V0 is also used for the centrality estimate via a fit of the distribution of the total signal amplitudes in the framework of the Glauber model~\cite{Abelev:2013qoq}. The reconstruction of the primary collision vertex is carried out in the two layers of the Silicon Pixel Detector (SPD), the innermost part of the Inner Tracking System of the experiment~\cite{Aamodt:2010aa}, covering $|\eta|<2$ and $|\eta|<1.4$  respectively. Finally, rejection of non-hadronic \mbox{Xe--Xe} collisions is performed using the Zero Degree Calorimeters  (ZDC)~\cite{ALICE:2012aa}, which identifies electromagnetic interactions, while the V0 detects beam-gas collisions occurring outside the nominal interaction point region.

The data analyzed in this Letter are taken with a trigger formed by the  coincidence of the MB trigger signal and of at least one muon triggered in the muon spectrometer, with a $p_{\rm T}=0.5$ GeV/$c$ threshold. The definition of the trigger is less restrictive than the one usually adopted for \mbox{Pb--Pb} data taking (1 GeV/$c$ threshold and two detected muons), due to the much smaller instantaneous luminosity for \mbox{Xe--Xe} collisions. Standard selection criteria~\cite{Adam:2016rdg} are then applied to such events and to the muon candidates. In particular, it is required (i) that two opposite-sign tracks reconstructed in the tracking chambers of the muon spectrometer are matched to track segments in the trigger system, (ii) that both muons belonging to the pair (dimuon) have $-4<\eta_{\mu}<-2.5$, and (iii) that their transverse position $R_{\rm abs}$ at the end of the hadron absorber of the muon spectrometer satisfies the condition $17.6<R_{\rm abs}<89.5$ cm. Finally, the reconstructed dimuon should lay in the fiducial rapidity region of the muon spectrometer, $2.5<y<4$.

The nuclear modification factor $R_{\rm AA}$ for the collision system under study is defined, for the centrality interval $i$, as 
\begin{eqnarray}
R_{\rm{AA}}^{i} = \frac{N_{{\rm J}/\psi}^{i}}{{\rm BR}_{\rm{J/}\psi\rightarrow\mu^{+}\mu^{-}}  N_{\rm{MB}}^{i}  A \epsilon^{i} \langle T_{\rm{AA}}^{i} \rangle   \sigma_{\rm{J/}\psi}^{\rm{pp}}},
\label{eq:raa}
\end{eqnarray}

where $N_{\rm{J/}\psi}^{i}$ is the number of detected J/$\psi$ in the $i$-th centrality interval, ${\rm BR}_{\rm{J/}\psi\rightarrow\mu^{+}\mu^{-}} = (5.96 \pm 0.03)$\% is the branching ratio of the dimuon decay channel~\cite{Patrignani:2016xqp}, $N_{\rm{MB}}^{i}$ is the number of MB events corresponding to the analyzed triggered event sample, $A\epsilon^{i}$ is the product of the detector acceptance times the reconstruction efficiency, $\langle T_{\rm{AA}}^{i} \rangle$ is the average nuclear thickness function~\cite{ALICE-PUBLIC-2018-003}, and  $\sigma_{\rm{J/}\psi}^{\rm{pp}}$ is the inclusive J/$\psi$ cross section for pp collisions, at the same energy and in the same kinematic range as the \mbox{Xe--Xe} data. Results are given for the centrality interval 0--90\% and for the two sub-intervals 0--20\% and 20-90\%.

Except for the determination of $\sigma_{\rm{J/}\psi}^{\rm{pp}}$, the other 
quantities entering the definition of $R_{\rm AA}$ are evaluated following the same procedure used for the analysis of the \mbox{Pb--Pb} data sample and detailed in Ref.~\cite{Adam:2016rdg}. 

The extraction of $N_{{\rm J}/\psi}$ is performed with two different approaches. In the first, the raw opposite-sign dimuon invariant mass distribution is fitted with a superposition of resonance and background shapes~\cite{ALICE-PUBLIC-2015-006}, the former being tuned to Monte Carlo (MC) simulations and the latter corresponding to empirical functions. In the second, the background is estimated via a mixed-event invariant mass distribution, obtained from the collected sample of muon-triggered events and subtracted from the raw spectrum~\cite{Adam:2015isa}. The resulting distribution is then fitted with the sum of a resonance shape and a continuum function accounting for the small residual background component. Due to the low statistical significance of the present data sample, the width of the J/$\psi$ meson, which is usually kept as a free parameter in the invariant mass fits, is fixed to $\sigma_{\rm J/\psi} = 70$ MeV/$c^2$, corresponding to the value of this quantity obtained in previous analyses~\cite{Abelev:2013yxa,Adam:2016rdg,Acharya:2017hjh}. For each of the two approaches, several fits were performed varying the fit mass range, the signal and background shapes and the J/$\psi$ width by $\pm 1$ MeV/$c^2$.
The obtained value for the centrality interval 0--90\% is $N_{{\rm J}/\psi} = 241\pm 47 ({\rm stat.)}\pm 26 ({\rm syst.})$, where the central value and the statistical uncertainty correspond to the average of the fit results and to the average of the corresponding statistical uncertainties, respectively.
The systematic uncertainty is obtained as the root mean square of the distribution of the $N_{{\rm J}/\psi}$ values obtained with the various fits.
The corresponding values for the 0--20\% and 20--90\% centrality sub-intervals are $N_{{\rm J}/\psi} = 175\pm 42 ({\rm stat.)}\pm 23 ({\rm syst.})$ and $N_{{\rm J}/\psi} = 77\pm 20 ({\rm stat.)}\pm 7 ({\rm syst.})$, respectively.

\begin{figure}
\begin{center}
\includegraphics[width=0.48\linewidth]{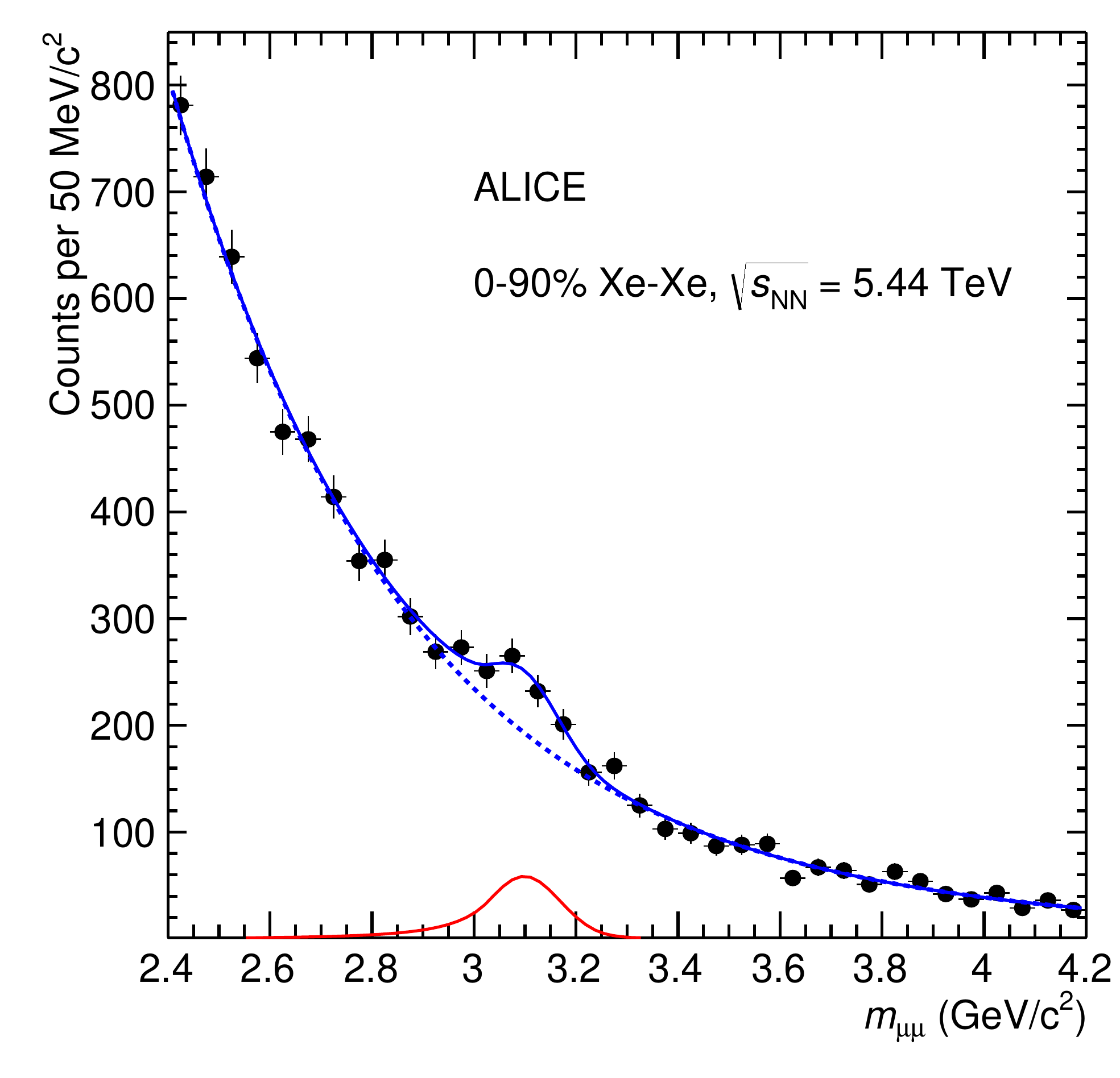}
\includegraphics[width=0.48\linewidth]{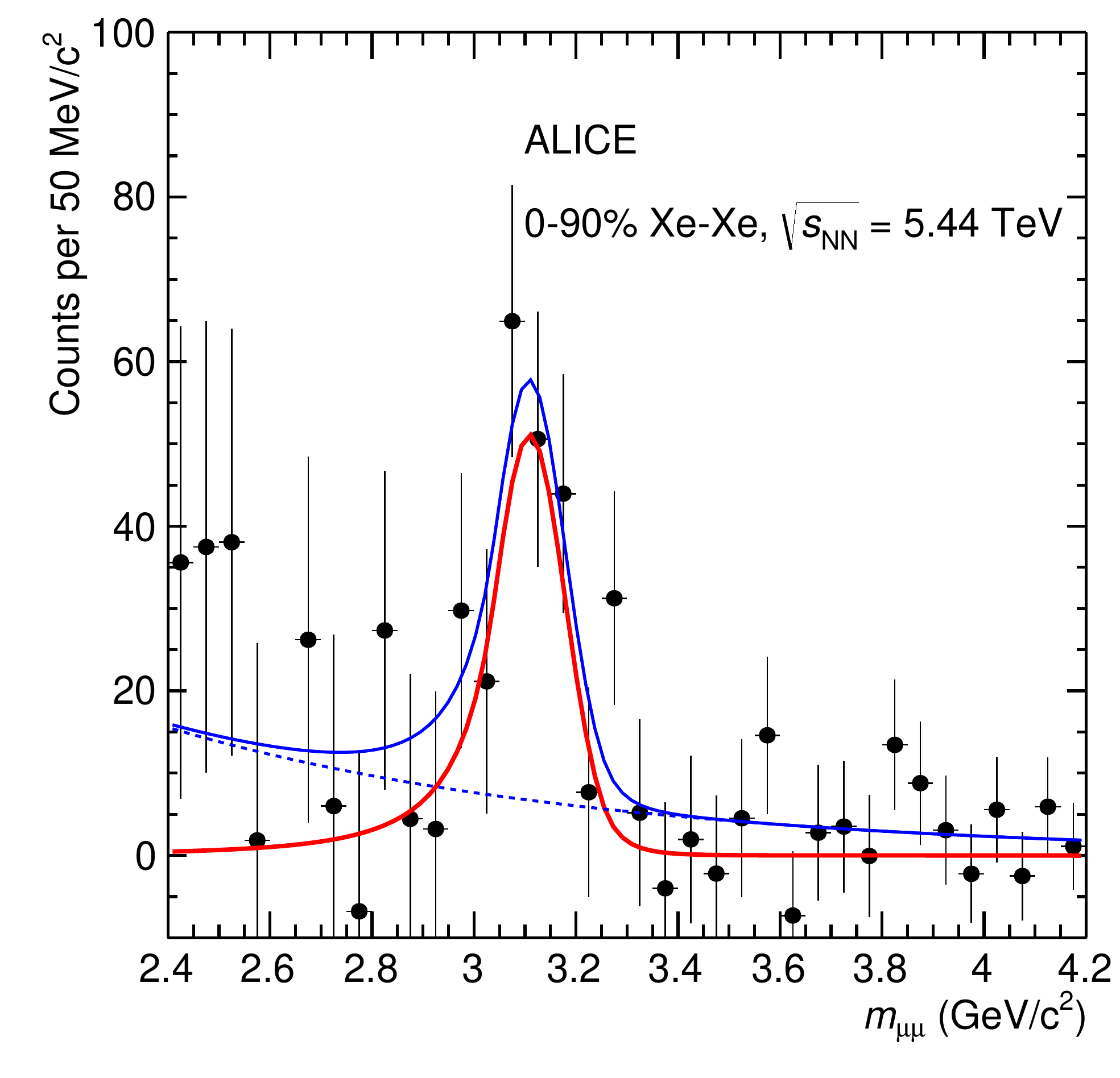}
\caption{Fits to invariant mass distributions of opposite-sign dimuons, for 0--90\% \mbox{Xe--Xe} collisions. 
In the left panel, the result of a fit to the raw invariant mass spectrum is shown, while in the right panel the fit to the same distribution after subtraction of the mixed-event background is presented. The fit curves shown in blue represent the sum of the signal and background shapes, while the red lines correspond to the J/$\psi$ signal and the blue dashed ones to the background (see text for details).}
\label{fig:invmass}
\end{center}
\end{figure}

Figure~\ref{fig:invmass} shows as an example the results of two fits to the 0--90\% Xe-Xe dimuon invariant mass distribution, corresponding to fitting the raw spectrum (left panel) or the mixed-event background subtracted mass distribution (right panel).

The product of the acceptance times the reconstruction efficiency $A\epsilon$ for J/$\psi$ is evaluated via a MC simulation, based on the GEANT3 transport model~\cite{Brun:1994aa}, which takes into account the alignment of the muon spectrometer detectors and their efficiency. 
The input $p_{\rm T}$ and $y$ distributions for the J/$\psi$ acceptance calculation cannot be tuned directly to data, due to the low integrated luminosity of the data sample. It is therefore assumed that the shape of the $y$ and $p_{\rm T}$ distributions is similar for different collision systems in centrality intervals corresponding to the same average number of participant nucleons,  weighted by the corresponding number of nucleon-nucleon collisions, $\langle N^{\rm w}_{\rm part}\rangle$ . The weighting is introduced to take into account that the J/$\psi$ production cross section is proportional to the number of nucleon-nucleon collisions and that therefore the average $N_{\rm part}$ in wide centrality bins is systematically shifted towards higher values. Following this argument, the differential distributions measured in \mbox{Pb--Pb} collisions at $\sqrt{s_{\rm NN}}=5.02$ TeV~\cite{Adam:2016rdg} for the 20--40\% centrality range are used as input distribution for the MC calculation, since  $\langle N^{\rm w}_{\rm part}\rangle_{\rm PbPb, 20-40\%}$ is equal, within $\sim 2$\%,  to $\langle N^{\rm w}_{\rm part}\rangle_{\rm XeXe, 0-90\%}$, estimated via a Glauber MC calculation. The systematic uncertainty on the J/$\psi$ acceptance value due to the choice of the J/$\psi$ rapidity and transverse momentum distributions amounts to 2\% and is evaluated by choosing alternative input shapes corresponding to other \mbox{Pb--Pb} centrality ranges.

Concerning the reconstruction efficiency, it slightly depends on the collision centrality, due to the detector occupancy in the muon spectrometer. The effect was evaluated in the analysis of \mbox{Pb--Pb} events~\cite{Adam:2016rdg} by embedding the simulated J/$\psi$ signal into real events corresponding to various centralities. For this analysis, starting from the \mbox{Pb--Pb} results, the decrease in $A\epsilon_{\rm XeXe, 0-90\%}$ with respect to a simulation containing only J/$\psi$ is estimated to be 4.2\% (values for 0--20\% and 20--90\% centrality ranges are 5.5\% and 1.6\%, respectively). The systematic uncertainty on the reconstruction efficiency is evaluated following the procedure used in Ref.~\cite{Adam:2016rdg}, leading to a 3.6\% effect.

The resulting value for the product of acceptance times reconstruction efficiency for J/$\psi$ production in 0--90\% Xe--Xe collisions is $A\epsilon_{\rm XeXe, 0-90\%}=0.228 \pm 0.009 ({\rm syst.})$, with a negligible statistical uncertainty.

The normalization factor $N_{\rm MB}$ is evaluated by multiplying the number of opposite-sign dimuon triggers by a factor $F_{\rm norm}$, corresponding to the inverse of the probability of having a triggered muon in a MB event. This quantity is computed from the event trigger input information and the level-0 trigger mask. The procedure and the evaluation of the systematic uncertainty are described in Ref.~\cite{Adam:2016rdg}. The obtained value is $F_{\rm norm}=2.428\pm 0.001 ({\rm stat.}) \pm 0.024 ({\rm syst.})$.

The reference cross section for the calculation of $R_{\rm AA}$ is obtained starting from the measured value of the inclusive J/$\psi$ cross section in pp collisions at $\sqrt{s}=5.02$ TeV~\cite{Adam:2016rdg}. This quantity is then corrected to account for the different centre-of-mass energy of the \mbox{Xe--Xe} data, using an interpolation of available ALICE pp results at $\sqrt{s}=$2.76, 5.02, 7, 8 and 13 TeV~\cite{Acharya:2017hjh}. The obtained value is $\sigma_{{\rm J}/\psi}^{\rm pp}=5.99 \pm 0.09 ({\rm stat.}) \pm 0.30 ({\rm syst.})$ $\mu$b$^{-1}$, where the systematic uncertainty contains a small term (0.4\%) related to the interpolation procedure, calculated as the maximum spread between results obtained with various interpolating functions~\cite{LHCb-CONF-2013-013}.

The nuclear thickness function $\langle T_{\rm AA}\rangle$ is evaluated for the various centrality intervals via a Glauber model calculation, and its uncertainty is estimated by varying within uncertainties the density parameters of the Xe nucleus~\cite{ALICE-PUBLIC-2018-003,Loizides:2017ack}. For 0--90\% centrality its value amounts to $\langle T_{\rm AA}\rangle = 3.25 \pm 0.25$ mb$^{-1}$, while for 0--20\% and 20--90\% one obtains $\langle T_{\rm AA}\rangle = 9.90 \pm 0.62$ mb$^{-1}$ and $\langle T_{\rm AA}\rangle = 1.35 \pm 0.14$ mb$^{-1}$, respectively.

\begin{table}
\begin{center}
\begin{tabular}{c|c|c|c}
\hline
Source & 0--90\% & 0--20\% & 20--90\% \\
\hline
Signal extraction & 11\% & 13\% & 8\% \\
\hline
MC input & 2\% & 2\% & 2\% \\
\hline
Tracking efficiency & 2\% & 2\% & 2\% \\
\hline
Trigger efficiency & 3\% & 3\% & 3\% \\
\hline
$F_{\rm norm}$ & 1\% & 1\% & 1\% \\
\hline
$\langle T_{\rm AA}\rangle$ & 8\% & 6\% & 10\% \\
\hline
Centrality & 0\% & 0\% & 1\% \\
\hline
pp reference & 5\% & 5\% & 5\% \\
\hline

\end{tabular}
\caption{Summary of systematic uncertainties on the calculation of the nuclear modification factors. The tracking efficiency term includes a 1\% contribution due to the choice of the $\chi^2$ cut of the matching between the information of tracking and trigger detectors. All the uncertainties are correlated among the various centrality ranges, except those on the signal extraction, $\langle T_{\rm AA}\rangle$ and the definition of the centrality intervals.}
\end{center}
\label{tab:syst}
\end{table}

\begin{figure}[h!]
\begin{center}
\includegraphics[width=0.8\linewidth]{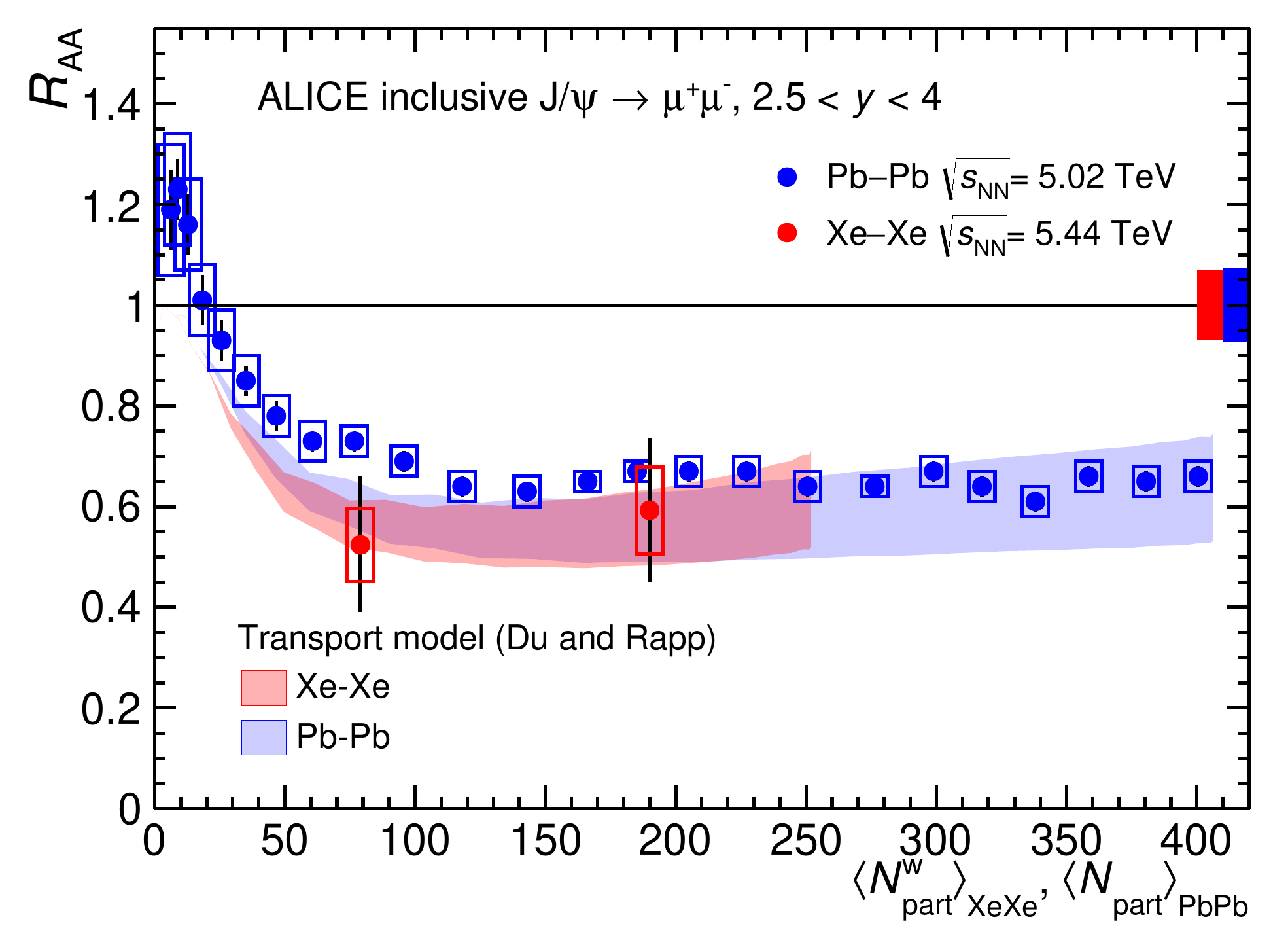}
\caption{The inclusive J/$\psi$ nuclear modification factor for \mbox{Xe--Xe} collisions at $\sqrt{s_{\rm NN}}=5.44$ TeV. The results are plotted using as centrality variable $\langle N^{\rm w}_{\rm part}\rangle$, obtained by weighting, in each centrality interval, the $N_{\rm part}$ distribution with the corresponding distribution of the number of nucleon-nucleon collisions. 
The error bars represent the statistical uncertainties, the boxes around the points the uncorrelated systematic uncertainties. Correlated uncertainties are shown as a filled box around unity.
The results are compared with the same quantity for \mbox{Pb--Pb} collisions at $\sqrt{s_{\rm NN}}=5.02$ TeV~\cite{Adam:2016rdg} and to the results of the calculation of a transport model~\cite{Zhao:2011cv,Du:2015wha}. 
For \mbox{Pb--Pb}, the weighting of $N_{\rm part}$ with the number of nucleon-nucleon collisions was not performed, since it leads to a negligible effect when the centrality intervals are narrow.}
\label{fig:RAAvscent}
\end{center}
\end{figure}

Finally, a systematic uncertainty on the definition of the centrality intervals is evaluated by varying the value of the V0 signal amplitude corresponding to 90\% centrality by $\pm 0.5$\% and recalculating correspondingly the centrality intervals.

Table 1 shows a summary of the systematic uncertainties for the $R_{\rm AA}$ measurement for the three analyzed centrality ranges. The main contributions come from the estimate of $\langle T_{\rm AA}\rangle$ and from the signal extraction. The former is dominated by the uncertainty on the surface thickness of the Xe nucleus. The latter, being estimated in a data-driven way as detailed above, may suffer from the statistical limitations of the data sample. The quoted values can therefore be considered to be a conservative estimate. 

The $p_{\rm T}$-integrated nuclear modification factor for inclusive J/$\psi$ production in \mbox{Xe--Xe} collisions at $\sqrt{s_{\rm NN}}=5.44$ TeV, measured in $2.5<y<4$ and in the 0--90\% centrality range, is $R_{\rm AA}=0.54 \pm 0.11 ({\rm stat.}) \pm 0.08 ({\rm syst.})$. This value can be compared with the corresponding one for \mbox{Pb--Pb} collisions at $\sqrt{s_{\rm NN}}=5.02$ TeV, $R_{\rm AA}^{\rm PbPb}=0.65 \pm 0.01 ({\rm stat.}) \pm 0.04 ({\rm syst.})$~\cite{Adam:2016rdg}.  Their ratio amounts to $0.84 \pm 0.16 ({\rm stat.}) \pm 0.13 ({\rm syst.})$, showing that the two values agree within about 0.8$\sigma$. Following the approach of Ref.~\cite{Adam:2015isa}, it can be shown that the Xe--Xe nuclear modification factor for prompt J/$\psi$ could be up to 10\% higher (lower) than the inclusive $R_{\rm AA}$ if the non-prompt J/$\psi$ component from the decays of hadrons containing a b quark is not (completely) suppressed.  
In Fig.~\ref{fig:RAAvscent} the $R_{\rm AA}$ values for 0--20\% and 20--90\% \mbox{Xe--Xe} collisions are plotted, and compared with the centrality dependence of the nuclear modification factor for \mbox{Pb--Pb} collisions~\cite{Adam:2016rdg}. The latter shows, after a decrease up to $N_{\rm part}\sim 100$, a saturation at $R_{\rm AA}\sim 0.65-0.7$ towards more central events, and the two \mbox{Xe--Xe} points are found to be in agreement, within their larger uncertainties, with the \mbox{Pb--Pb} results. The \mbox{Xe--Xe} and \mbox{Pb--Pb} results are also compared with the calculation of a transport model by Du and Rapp~\cite{Zhao:2011cv,Du:2015wha}. 
A close similarity of the predicted suppression patterns for \mbox{Pb--Pb} and \mbox{Xe--Xe} is observed, which fairly reproduces the experimental results.

In summary, we have measured inclusive J/$\psi$ production in \mbox{Xe--Xe} collisions at $\sqrt{s_{\rm NN}}=5.44$ TeV. Results on the nuclear modification factors were given for various centrality selections and compared to corresponding results for \mbox{Pb--Pb} collisions at $\sqrt{s_{\rm NN}}=5.02$ TeV and to a theoretical model. Within the experimental uncertainties, a good agreement is found between the $R_{\rm AA}$ measured in the two systems and with the calculation. These results show that the relative contribution of suppression and regeneration processes is similar for collisions producing similar $N_{\rm part}$ values from different collision systems.
%
%

\newenvironment{acknowledgement}{\relax}{\relax}
\begin{acknowledgement}
\section*{Acknowledgements}

The ALICE Collaboration would like to thank all its engineers and technicians for their invaluable contributions to the construction of the experiment and the CERN accelerator teams for the outstanding performance of the LHC complex.
The ALICE Collaboration gratefully acknowledges the resources and support provided by all Grid centres and the Worldwide LHC Computing Grid (WLCG) collaboration.
The ALICE Collaboration acknowledges the following funding agencies for their support in building and running the ALICE detector:
A. I. Alikhanyan National Science Laboratory (Yerevan Physics Institute) Foundation (ANSL), State Committee of Science and World Federation of Scientists (WFS), Armenia;
Austrian Academy of Sciences and Nationalstiftung f\"{u}r Forschung, Technologie und Entwicklung, Austria;
Ministry of Communications and High Technologies, National Nuclear Research Center, Azerbaijan;
Conselho Nacional de Desenvolvimento Cient\'{\i}fico e Tecnol\'{o}gico (CNPq), Universidade Federal do Rio Grande do Sul (UFRGS), Financiadora de Estudos e Projetos (Finep) and Funda\c{c}\~{a}o de Amparo \`{a} Pesquisa do Estado de S\~{a}o Paulo (FAPESP), Brazil;
Ministry of Science \& Technology of China (MSTC), National Natural Science Foundation of China (NSFC) and Ministry of Education of China (MOEC) , China;
Ministry of Science and Education, Croatia;
Ministry of Education, Youth and Sports of the Czech Republic, Czech Republic;
The Danish Council for Independent Research | Natural Sciences, the Carlsberg Foundation and Danish National Research Foundation (DNRF), Denmark;
Helsinki Institute of Physics (HIP), Finland;
Commissariat \`{a} l'Energie Atomique (CEA) and Institut National de Physique Nucl\'{e}aire et de Physique des Particules (IN2P3) and Centre National de la Recherche Scientifique (CNRS), France;
Bundesministerium f\"{u}r Bildung, Wissenschaft, Forschung und Technologie (BMBF) and GSI Helmholtzzentrum f\"{u}r Schwerionenforschung GmbH, Germany;
General Secretariat for Research and Technology, Ministry of Education, Research and Religions, Greece;
National Research, Development and Innovation Office, Hungary;
Department of Atomic Energy Government of India (DAE), Department of Science and Technology, Government of India (DST), University Grants Commission, Government of India (UGC) and Council of Scientific and Industrial Research (CSIR), India;
Indonesian Institute of Science, Indonesia;
Centro Fermi - Museo Storico della Fisica e Centro Studi e Ricerche Enrico Fermi and Istituto Nazionale di Fisica Nucleare (INFN), Italy;
Institute for Innovative Science and Technology , Nagasaki Institute of Applied Science (IIST), Japan Society for the Promotion of Science (JSPS) KAKENHI and Japanese Ministry of Education, Culture, Sports, Science and Technology (MEXT), Japan;
Consejo Nacional de Ciencia (CONACYT) y Tecnolog\'{i}a, through Fondo de Cooperaci\'{o}n Internacional en Ciencia y Tecnolog\'{i}a (FONCICYT) and Direcci\'{o}n General de Asuntos del Personal Academico (DGAPA), Mexico;
Nederlandse Organisatie voor Wetenschappelijk Onderzoek (NWO), Netherlands;
The Research Council of Norway, Norway;
Commission on Science and Technology for Sustainable Development in the South (COMSATS), Pakistan;
Pontificia Universidad Cat\'{o}lica del Per\'{u}, Peru;
Ministry of Science and Higher Education and National Science Centre, Poland;
Korea Institute of Science and Technology Information and National Research Foundation of Korea (NRF), Republic of Korea;
Ministry of Education and Scientific Research, Institute of Atomic Physics and Romanian National Agency for Science, Technology and Innovation, Romania;
Joint Institute for Nuclear Research (JINR), Ministry of Education and Science of the Russian Federation and National Research Centre Kurchatov Institute, Russia;
Ministry of Education, Science, Research and Sport of the Slovak Republic, Slovakia;
National Research Foundation of South Africa, South Africa;
Centro de Aplicaciones Tecnol\'{o}gicas y Desarrollo Nuclear (CEADEN), Cubaenerg\'{\i}a, Cuba and Centro de Investigaciones Energ\'{e}ticas, Medioambientales y Tecnol\'{o}gicas (CIEMAT), Spain;
Swedish Research Council (VR) and Knut \& Alice Wallenberg Foundation (KAW), Sweden;
European Organization for Nuclear Research, Switzerland;
National Science and Technology Development Agency (NSDTA), Suranaree University of Technology (SUT) and Office of the Higher Education Commission under NRU project of Thailand, Thailand;
Turkish Atomic Energy Agency (TAEK), Turkey;
National Academy of  Sciences of Ukraine, Ukraine;
Science and Technology Facilities Council (STFC), United Kingdom;
National Science Foundation of the United States of America (NSF) and United States Department of Energy, Office of Nuclear Physics (DOE NP), United States of America.
\end{acknowledgement}

\bibliographystyle{utphys}   
\bibliography{JpsiXeXe_paper_v6}

\newpage
\appendix
\section{The ALICE Collaboration}
\label{app:collab}

\begingroup
\small
\begin{flushleft}
S.~Acharya\Irefn{org139}\And 
F.T.-.~Acosta\Irefn{org22}\And 
D.~Adamov\'{a}\Irefn{org94}\And 
J.~Adolfsson\Irefn{org81}\And 
M.M.~Aggarwal\Irefn{org98}\And 
G.~Aglieri Rinella\Irefn{org36}\And 
M.~Agnello\Irefn{org33}\And 
N.~Agrawal\Irefn{org49}\And 
Z.~Ahammed\Irefn{org139}\And 
S.U.~Ahn\Irefn{org77}\And 
S.~Aiola\Irefn{org144}
% ...
\And
\renewcommand\labelenumi{\textsuperscript{\theenumi}~}

\section*{Affiliation notes}
\renewcommand\theenumi{\roman{enumi}}
\begin{Authlist}
\item \Adef{org*}Deceased
\item \Adef{orgI}Dipartimento DET del Politecnico di Torino, Turin, Italy
\item \Adef{orgII}M.V. Lomonosov Moscow State University, D.V. Skobeltsyn Institute of Nuclear, Physics, Moscow, Russia
\item \Adef{orgIII}Department of Applied Physics, Aligarh Muslim University, Aligarh, India
\item \Adef{orgIV}Institute of Theoretical Physics, University of Wroclaw, Poland
\end{Authlist}

\section*{Collaboration Institutes}
\renewcommand\theenumi{\arabic{enumi}~}
\begin{Authlist}
\item \Idef{org1}A.I. Alikhanyan National Science Laboratory (Yerevan Physics Institute) Foundation, Yerevan, Armenia
\item \Idef{org2}Benem\'{e}rita Universidad Aut\'{o}noma de Puebla, Puebla, Mexico
\item \Idef{org3}Bogolyubov Institute for Theoretical Physics, National Academy of Sciences of Ukraine, Kiev, Ukraine
\item \Idef{org4}Bose Institute, Department of Physics  and Centre for Astroparticle Physics and Space Science (CAPSS), Kolkata, India
\item \Idef{org5}Budker Institute for Nuclear Physics, Novosibirsk, Russia
\item \Idef{org6}California Polytechnic State University, San Luis Obispo, California, United States
\item \Idef{org7}Central China Normal University, Wuhan, China
\item \Idef{org8}Centre de Calcul de l'IN2P3, Villeurbanne, Lyon, France
\item \Idef{org9}Centro de Aplicaciones Tecnol\'{o}gicas y Desarrollo Nuclear (CEADEN), Havana, Cuba
\item \Idef{org10}Centro de Investigaci\'{o}n y de Estudios Avanzados (CINVESTAV), Mexico City and M\'{e}rida, Mexico
\item \Idef{org11}Centro Fermi - Museo Storico della Fisica e Centro Studi e Ricerche ``Enrico Fermi', Rome, Italy
\item \Idef{org12}Chicago State University, Chicago, Illinois, United States
\item \Idef{org13}China Institute of Atomic Energy, Beijing, China
\item \Idef{org14}Chonbuk National University, Jeonju, Republic of Korea
\item \Idef{org15}Comenius University Bratislava, Faculty of Mathematics, Physics and Informatics, Bratislava, Slovakia
\item \Idef{org16}COMSATS Institute of Information Technology (CIIT), Islamabad, Pakistan
\item \Idef{org17}Creighton University, Omaha, Nebraska, United States
\item \Idef{org18}Department of Physics, Aligarh Muslim University, Aligarh, India
\item \Idef{org19}Department of Physics, Ohio State University, Columbus, Ohio, United States
\item \Idef{org20}Department of Physics, Pusan National University, Pusan, Republic of Korea
\item \Idef{org21}Department of Physics, Sejong University, Seoul, Republic of Korea
\item \Idef{org22}Department of Physics, University of California, Berkeley, California, United States
\item \Idef{org23}Department of Physics, University of Oslo, Oslo, Norway
\item \Idef{org24}Department of Physics and Technology, University of Bergen, Bergen, Norway
\item \Idef{org25}Dipartimento di Fisica dell'Universit\`{a} 'La Sapienza' and Sezione INFN, Rome, Italy
\item \Idef{org26}Dipartimento di Fisica dell'Universit\`{a} and Sezione INFN, Cagliari, Italy
\item \Idef{org27}Dipartimento di Fisica dell'Universit\`{a} and Sezione INFN, Trieste, Italy
\item \Idef{org28}Dipartimento di Fisica dell'Universit\`{a} and Sezione INFN, Turin, Italy
\item \Idef{org29}Dipartimento di Fisica e Astronomia dell'Universit\`{a} and Sezione INFN, Bologna, Italy
\item \Idef{org30}Dipartimento di Fisica e Astronomia dell'Universit\`{a} and Sezione INFN, Catania, Italy
\item \Idef{org31}Dipartimento di Fisica e Astronomia dell'Universit\`{a} and Sezione INFN, Padova, Italy
\item \Idef{org32}Dipartimento di Fisica `E.R.~Caianiello' dell'Universit\`{a} and Gruppo Collegato INFN, Salerno, Italy
\item \Idef{org33}Dipartimento DISAT del Politecnico and Sezione INFN, Turin, Italy
\item \Idef{org34}Dipartimento di Scienze e Innovazione Tecnologica dell'Universit\`{a} del Piemonte Orientale and INFN Sezione di Torino, Alessandria, Italy
\item \Idef{org35}Dipartimento Interateneo di Fisica `M.~Merlin' and Sezione INFN, Bari, Italy
\item \Idef{org36}European Organization for Nuclear Research (CERN), Geneva, Switzerland
\item \Idef{org37}Faculty of Electrical Engineering, Mechanical Engineering and Naval Architecture, University of Split, Split, Croatia
\item \Idef{org38}Faculty of Engineering and Science, Western Norway University of Applied Sciences, Bergen, Norway
\item \Idef{org39}Faculty of Nuclear Sciences and Physical Engineering, Czech Technical University in Prague, Prague, Czech Republic
\item \Idef{org40}Faculty of Science, P.J.~\v{S}af\'{a}rik University, Ko\v{s}ice, Slovakia
\item \Idef{org41}Frankfurt Institute for Advanced Studies, Johann Wolfgang Goethe-Universit\"{a}t Frankfurt, Frankfurt, Germany
\item \Idef{org42}Gangneung-Wonju National University, Gangneung, Republic of Korea
\item \Idef{org43}Gauhati University, Department of Physics, Guwahati, India
\item \Idef{org44}Helmholtz-Institut f\"{u}r Strahlen- und Kernphysik, Rheinische Friedrich-Wilhelms-Universit\"{a}t Bonn, Bonn, Germany
\item \Idef{org45}Helsinki Institute of Physics (HIP), Helsinki, Finland
\item \Idef{org46}Hiroshima University, Hiroshima, Japan
\item \Idef{org47}Hochschule Worms, Zentrum  f\"{u}r Technologietransfer und Telekommunikation (ZTT), Worms, Germany
\item \Idef{org48}Horia Hulubei National Institute of Physics and Nuclear Engineering, Bucharest, Romania
\item \Idef{org49}Indian Institute of Technology Bombay (IIT), Mumbai, India
\item \Idef{org50}Indian Institute of Technology Indore, Indore, India
\item \Idef{org51}Indonesian Institute of Sciences, Jakarta, Indonesia
\item \Idef{org52}INFN, Laboratori Nazionali di Frascati, Frascati, Italy
\item \Idef{org53}INFN, Sezione di Bari, Bari, Italy
\item \Idef{org54}INFN, Sezione di Bologna, Bologna, Italy
\item \Idef{org55}INFN, Sezione di Cagliari, Cagliari, Italy
\item \Idef{org56}INFN, Sezione di Catania, Catania, Italy
\item \Idef{org57}INFN, Sezione di Padova, Padova, Italy
\item \Idef{org58}INFN, Sezione di Roma, Rome, Italy
\item \Idef{org59}INFN, Sezione di Torino, Turin, Italy
\item \Idef{org60}INFN, Sezione di Trieste, Trieste, Italy
\item \Idef{org61}Inha University, Incheon, Republic of Korea
\item \Idef{org62}Institut de Physique Nucl\'{e}aire d'Orsay (IPNO), Institut National de Physique Nucl\'{e}aire et de Physique des Particules (IN2P3/CNRS), Universit\'{e} de Paris-Sud, Universit\'{e} Paris-Saclay, Orsay, France
\item \Idef{org63}Institute for Nuclear Research, Academy of Sciences, Moscow, Russia
\item \Idef{org64}Institute for Subatomic Physics, Utrecht University/Nikhef, Utrecht, Netherlands
\item \Idef{org65}Institute for Theoretical and Experimental Physics, Moscow, Russia
\item \Idef{org66}Institute of Experimental Physics, Slovak Academy of Sciences, Ko\v{s}ice, Slovakia
\item \Idef{org67}Institute of Physics, Bhubaneswar, India
\item \Idef{org68}Institute of Physics of the Czech Academy of Sciences, Prague, Czech Republic
\item \Idef{org69}Institute of Space Science (ISS), Bucharest, Romania
\item \Idef{org70}Institut f\"{u}r Kernphysik, Johann Wolfgang Goethe-Universit\"{a}t Frankfurt, Frankfurt, Germany
\item \Idef{org71}Instituto de Ciencias Nucleares, Universidad Nacional Aut\'{o}noma de M\'{e}xico, Mexico City, Mexico
\item \Idef{org72}Instituto de F\'{i}sica, Universidade Federal do Rio Grande do Sul (UFRGS), Porto Alegre, Brazil
\item \Idef{org73}Instituto de F\'{\i}sica, Universidad Nacional Aut\'{o}noma de M\'{e}xico, Mexico City, Mexico
\item \Idef{org74}iThemba LABS, National Research Foundation, Somerset West, South Africa
\item \Idef{org75}Johann-Wolfgang-Goethe Universit\"{a}t Frankfurt Institut f\"{u}r Informatik, Fachbereich Informatik und Mathematik, Frankfurt, Germany
\item \Idef{org76}Joint Institute for Nuclear Research (JINR), Dubna, Russia
\item \Idef{org77}Korea Institute of Science and Technology Information, Daejeon, Republic of Korea
\item \Idef{org78}KTO Karatay University, Konya, Turkey
\item \Idef{org79}Laboratoire de Physique Subatomique et de Cosmologie, Universit\'{e} Grenoble-Alpes, CNRS-IN2P3, Grenoble, France
\item \Idef{org80}Lawrence Berkeley National Laboratory, Berkeley, California, United States
\item \Idef{org81}Lund University Department of Physics, Division of Particle Physics, Lund, Sweden
\item \Idef{org82}Nagasaki Institute of Applied Science, Nagasaki, Japan
\item \Idef{org83}Nara Women{'}s University (NWU), Nara, Japan
\item \Idef{org84}National and Kapodistrian University of Athens, School of Science, Department of Physics , Athens, Greece
\item \Idef{org85}National Centre for Nuclear Research, Warsaw, Poland
\item \Idef{org86}National Institute of Science Education and Research, HBNI, Jatni, India
\item \Idef{org87}National Nuclear Research Center, Baku, Azerbaijan
\item \Idef{org88}National Research Centre Kurchatov Institute, Moscow, Russia
\item \Idef{org89}Niels Bohr Institute, University of Copenhagen, Copenhagen, Denmark
\item \Idef{org90}Nikhef, National institute for subatomic physics, Amsterdam, Netherlands
\item \Idef{org91}NRC Kurchatov Institute IHEP, Protvino, Russia
\item \Idef{org92}NRNU Moscow Engineering Physics Institute, Moscow, Russia
\item \Idef{org93}Nuclear Physics Group, STFC Daresbury Laboratory, Daresbury, United Kingdom
\item \Idef{org94}Nuclear Physics Institute of the Czech Academy of Sciences, \v{R}e\v{z} u Prahy, Czech Republic
\item \Idef{org95}Oak Ridge National Laboratory, Oak Ridge, Tennessee, United States
\item \Idef{org96}Petersburg Nuclear Physics Institute, Gatchina, Russia
\item \Idef{org97}Physics department, Faculty of science, University of Zagreb, Zagreb, Croatia
\item \Idef{org98}Physics Department, Panjab University, Chandigarh, India
\item \Idef{org99}Physics Department, University of Jammu, Jammu, India
\item \Idef{org100}Physics Department, University of Rajasthan, Jaipur, India
\item \Idef{org101}Physikalisches Institut, Eberhard-Karls-Universit\"{a}t T\"{u}bingen, T\"{u}bingen, Germany
\item \Idef{org102}Physikalisches Institut, Ruprecht-Karls-Universit\"{a}t Heidelberg, Heidelberg, Germany
\item \Idef{org103}Physik Department, Technische Universit\"{a}t M\"{u}nchen, Munich, Germany
\item \Idef{org104}Research Division and ExtreMe Matter Institute EMMI, GSI Helmholtzzentrum f\"ur Schwerionenforschung GmbH, Darmstadt, Germany
\item \Idef{org105}Rudjer Bo\v{s}kovi\'{c} Institute, Zagreb, Croatia
\item \Idef{org106}Russian Federal Nuclear Center (VNIIEF), Sarov, Russia
\item \Idef{org107}Saha Institute of Nuclear Physics, Kolkata, India
\item \Idef{org108}School of Physics and Astronomy, University of Birmingham, Birmingham, United Kingdom
\item \Idef{org109}Secci\'{o}n F\'{\i}sica, Departamento de Ciencias, Pontificia Universidad Cat\'{o}lica del Per\'{u}, Lima, Peru
\item \Idef{org110}Shanghai Institute of Applied Physics, Shanghai, China
\item \Idef{org111}Stefan Meyer Institut f\"{u}r Subatomare Physik (SMI), Vienna, Austria
\item \Idef{org112}SUBATECH, IMT Atlantique, Universit\'{e} de Nantes, CNRS-IN2P3, Nantes, France
\item \Idef{org113}Suranaree University of Technology, Nakhon Ratchasima, Thailand
\item \Idef{org114}Technical University of Ko\v{s}ice, Ko\v{s}ice, Slovakia
\item \Idef{org115}Technische Universit\"{a}t M\"{u}nchen, Excellence Cluster 'Universe', Munich, Germany
\item \Idef{org116}The Henryk Niewodniczanski Institute of Nuclear Physics, Polish Academy of Sciences, Cracow, Poland
\item \Idef{org117}The University of Texas at Austin, Austin, Texas, United States
\item \Idef{org118}Universidad Aut\'{o}noma de Sinaloa, Culiac\'{a}n, Mexico
\item \Idef{org119}Universidade de S\~{a}o Paulo (USP), S\~{a}o Paulo, Brazil
\item \Idef{org120}Universidade Estadual de Campinas (UNICAMP), Campinas, Brazil
\item \Idef{org121}Universidade Federal do ABC, Santo Andre, Brazil
\item \Idef{org122}University College of Southeast Norway, Tonsberg, Norway
\item \Idef{org123}University of Cape Town, Cape Town, South Africa
\item \Idef{org124}University of Houston, Houston, Texas, United States
\item \Idef{org125}University of Jyv\"{a}skyl\"{a}, Jyv\"{a}skyl\"{a}, Finland
\item \Idef{org126}University of Liverpool, Department of Physics Oliver Lodge Laboratory , Liverpool, United Kingdom
\item \Idef{org127}University of Tennessee, Knoxville, Tennessee, United States
\item \Idef{org128}University of the Witwatersrand, Johannesburg, South Africa
\item \Idef{org129}University of Tokyo, Tokyo, Japan
\item \Idef{org130}University of Tsukuba, Tsukuba, Japan
\item \Idef{org131}Universit\'{e} Clermont Auvergne, CNRS/IN2P3, LPC, Clermont-Ferrand, France
\item \Idef{org132}Universit\'{e} de Lyon, Universit\'{e} Lyon 1, CNRS/IN2P3, IPN-Lyon, Villeurbanne, Lyon, France
\item \Idef{org133}Universit\'{e} de Strasbourg, CNRS, IPHC UMR 7178, F-67000 Strasbourg, France, Strasbourg, France
\item \Idef{org134} Universit\'{e} Paris-Saclay Centre d¿\'Etudes de Saclay (CEA), IRFU, Department de Physique Nucl\'{e}aire (DPhN), Saclay, France
\item \Idef{org135}Universit\`{a} degli Studi di Foggia, Foggia, Italy
\item \Idef{org136}Universit\`{a} degli Studi di Pavia, Pavia, Italy
\item \Idef{org137}Universit\`{a} di Brescia, Brescia, Italy
\item \Idef{org138}V.~Fock Institute for Physics, St. Petersburg State University, St. Petersburg, Russia
\item \Idef{org139}Variable Energy Cyclotron Centre, Kolkata, India
\item \Idef{org140}Warsaw University of Technology, Warsaw, Poland
\item \Idef{org141}Wayne State University, Detroit, Michigan, United States
\item \Idef{org142}Westf\"{a}lische Wilhelms-Universit\"{a}t M\"{u}nster, Institut f\"{u}r Kernphysik, M\"{u}nster, Germany
\item \Idef{org143}Wigner Research Centre for Physics, Hungarian Academy of Sciences, Budapest, Hungary
\item \Idef{org144}Yale University, New Haven, Connecticut, United States
\item \Idef{org145}Yonsei University, Seoul, Republic of Korea
\end{Authlist}
\endgroup
\end{document}